\documentclass[aps,prl,twocolumn,footinbib,superscriptaddress,floatfix,showpacs,amsmath,amsfonts,amssymb]{revtex4-2}
\usepackage{physics}
\usepackage{times}
\usepackage{float}
\usepackage{multirow}
\usepackage[dvipsnames]{xcolor} 
\usepackage{amsbsy}
\usepackage{enumitem}
\usepackage{wasysym}
\usepackage[english]{babel}
\usepackage[T1]{fontenc}
\usepackage[utf8]{inputenc} 
\usepackage{graphicx}
\usepackage[colorlinks,bookmarks=false,citecolor=blue,linkcolor=red,urlcolor=blue]{hyperref}
\usepackage{pstricks}
\usepackage{rotating}			       
\usepackage{tabularx,hhline}	
\usepackage[caption=false]{subfig}		
\newcolumntype{P}[1]{>{\centering\arraybackslash}p{#1}}
\usepackage{soul}
\usepackage[caption=false]{subfig}

\usepackage{amsmath}

\usepackage{pdfpages}

\makeatletter
\AtBeginDocument{\let\LS@rot\@undefined}
\makeatother

\colorlet{mylinkcolor}{violet}
\colorlet{mycitecolor}{YellowOrange}
\colorlet{myurlcolor}{Aquamarine}

\renewcommand\[{\begin{equation}}
\renewcommand\]{\end{equation}}
\begin{document}

\title{Gravitational wave analogues in spin nematics and cold atoms}

\author{Leilee Chojnacki}
\email{leilee.chojnacki@oist.jp}
\affiliation{Theory of Quantum Matter Unit, Okinawa Institute of Science and Technology Graduate University, Onna-son, Okinawa 904-0412, Japan} 

\author{Rico Pohle}
\affiliation{Department of Applied Physics, The University of Tokyo, Hongo,  Bunkyo-ku,
Tokyo, 113-8656, Japan}
\affiliation{Graduate School of Science and Technology, Keio University, 
Yokohama 223-8522, Japan}

\author{Han Yan}
\affiliation{Department of Physics \& Astronomy, Rice University, Houston, TX 77005, USA}
\affiliation{Smalley-Curl Institute, Rice University, Houston, TX 77005, USA}
\affiliation{Institute for Solid State Physics, University of Tokyo, Kashiwa, 277-8581 Chiba, Japan}

\author{Yutaka Akagi}
\affiliation{Department of Physics, Graduate School of Science, The University of Tokyo, Hongo, Tokyo 113-0033, Japan}

\author{Nic Shannon}
\affiliation{Theory of Quantum Matter Unit, Okinawa Institute of Science and Technology Graduate University, Onna-son, Okinawa 904-0412, Japan}

\date{\today}

\begin{abstract}

Many large-scale phenomena in our Universe, such as gravitational waves, are challenging to reproduce in laboratory settings.  
However, parallels with condensed matter systems can provide alternative routes for experimental accessibility.  
Here we show how spin nematic phases provide a low-energy avenue for 
accessing the physics of linearized gravity, and in particular that their Goldstone modes are relativistically-dispersing massless spin-2 excitations, analogous to gravitational waves.  
We show at the level of the action that the low-energy effective field theory describing a spin nematic is in correspondence with that of linearized gravity.  
We then explicitly identify a microscopic model of a spin-1 magnet whose excitations in the low energy limit are relativistically dispersing, massless spin-2 Bosons which are in one-to-one correspondence with gravitational waves and, supported by simulation, outline a procedure 
for directly observing these analogue waves in a cold gas of $^{23}$Na atoms.

\end{abstract}


\maketitle

\textit{Introduction.--} 
Light has been a natural companion of humanity since our earliest days, shaping civilization as we know it. 
However, our attention to astrophysical gravitational waves is, by comparison, still in its infancy. 
The experimental detection of gravitational waves by the LIGO collaboration \cite{Abbott2016PhysRevLett} 
marked the beginning of a new age of observational astronomy. 
That said, production of measurable gravitational radiation is far from being feasible due 
to the energy scales involved, unlike its photonic counterpart. 
Alternatives that provide laboratory access to such massless spin-2 waves 
would therefore provide many new opportunities.


Thus far, several condensed matter systems have been suggested to mimic features 
of gravity, with much focus on reproducing the effects of curved spacetimes. 
Acoustic analogues of gravitational phenomena were first suggested by Unruh \cite{unruh_sonic_1995} and later measured \cite{Weinfurtner2011PhysRevLett}, with many further promising experimental candidates in superfluids \cite{volovik_simulation_1998,Volovik2009}, in semimetals \cite{volovik_black_2016,liu_fermionic_2020}, in quantum Hall systems \cite{Yusa2022PhysRevD}, in optics \cite{rosenberg_optical_2020,petty_optical_2020} and in cold atoms \cite{bravo_analog_2015, sabin_thermal_2016,viermann_quantum_2022,tolosa-simeon_curved_2022}. 
 In the theory domain, connections between elasticity and emergent gravitational phenomena 
 have been studied by Kleinert \textit{et al.} \cite{kleinert_gravity_1987,kleinert_nematic_2004,zhu_world_2011,zaanen_emergence_2012,beekman_dual_2017,beekman_dual_2017-1} 
 and independently in the context of fracton models \cite{pretko_fracton-elasticity_2018,pretko_emergent_2017,Yan2019PhysRevBfracton1,Yan2019PhysRevBfracton2,Yan2020PhysRevB,yan2023padic}, 
 and related aspects of geometry also arise in magnetic models \cite{hill_chiral_2021} and graphene \cite{roberts2023analog}.
However, no experimentally viable platform has yet been realized that provides direct access to massless spin-2 Bosons akin to gravitational waves, even in the flat spacetime analogue.


In this Letter we identify a parallel between gravitational waves and quadrupolar waves in 
quantum spin nematics, and suggest two routes for their experimental realization. 
We first review the description of gravitational waves within linearized gravity.
We then show that an identical set of equations arises in the low-energy continuum 
field theory describing spin nematics.
Through numerical simulation, we explore the real-time dynamics of a microscopic model 
with spin nematic order, showing how quadrupolar waves---equivalent to gravitational 
waves---are generated through the annihilation of topological defects.
We conclude by suggesting an experimental protocol for the creation and observation of 
analogue gravitational waves in spin nematic phases, realized in either magnetic insulators 
or cold atoms.

%
\textit{Linearized gravity and gravitational waves.--} 
We now briefly summarize the key features of linearized gravity, leading up to gravitational waves.
This treatment follows the conventions of standard textbooks, e.g., Refs~\cite{Misner1973, Carroll2019, Maggiore2007}.
General relativity (GR) is a geometrical theory, describing the curvature of a 4-dimensional spacetime.
Fundamental to this is the metric tensor, $g_{\mu\nu}$, a symmetric rank-2 tensor, 
which allows the definition of distance. Here the Greek indices 
$\mu$, $\nu$ run over all four spacetime dimensions.
In linearized gravity, spacetime is assumed flat up to small fluctuations, $h_{\mu\nu}$, such that 
\begin{eqnarray}
g_{\mu\nu} =  \eta_{\mu\nu} + h_{\mu\nu},
\label{eq:metric.tensor}
\end{eqnarray}
where $\eta_{\mu\nu} = \text{diag} (-1,1,1,1)$ is the Minkowski metric for a flat spacetime.   
The linearized theory is invariant under transformations 
\begin{subequations}
\begin{eqnarray}
x^{\prime\mu} &=& x^{\mu} + \xi^{\mu}(x) \; , \\
h_{\mu\nu}^{\prime} &=& h_{\mu\nu} - \partial_{\nu}\xi_{\mu} - \partial_{\mu}\xi_{\nu} \; ,
\end{eqnarray}
\end{subequations}
where $x^\mu$ denotes spacetime coordinates, and $\xi_{\mu}$ corresponds to an infinitesimal 
coordinate transformation.
The existence of these transformations implies that not all degrees of freedom are independent, and 
in deriving a theory for gravitational waves, it is conventional to make the choice 
\begin{subequations}
\begin{align}
	&h^\mu_{\ \mu}(x^\sigma) = 0   \;  [\textrm{traceless}] \; , \\
	&h_{0\mu}(x^\sigma)= 0  \; [\textrm{no scalar or vector components}]\; , \\ 
	&\partial^nh_{nm}(x^\sigma)= 0 \; \; ,  \nonumber\\ 
    \implies &k^nh_{nm}(k^\sigma) = 0  \;  [\textrm{no longitudinal dynamics}] \; .
    \label{eq:constraints.LG3}
\end{align} 
\end{subequations}
Here, Roman indices $n$, $m$ denote the spatial components, and the Einstein 
summation convention for repeated indices is assumed.
Implementing these constraints, we arrive at a theory expressed in terms 
of a symmetric, traceless, rank-2 tensor \cite{Maggiore2007}, with dynamics governed 
by the action
\begin{eqnarray}
	\mathcal{S}_\mathsf{LG} = -\frac{c^3}{16\pi G} \int d^4x\; \bigg[\partial^{\alpha}h^{\mu\nu}\partial_{\alpha}h_{\mu\nu}\bigg] \; ,
\label{eq:S.linearized.gravity}
\end{eqnarray}
where $c$ is the speed of light, and $G$ the gravitational constant.
This leads to the equation of motion for massless waves
\begin{eqnarray}
	\frac{1}{c^2}\partial_t\partial^t h_{\mu\nu} - \partial_n \partial^n h_{\mu\nu} = 0 \; , 
\label{eq:EoM.linearized.gravity}	
\end{eqnarray}
where implicitly, only two of the 16 components of $h_{\mu\nu}$ have non-trivial independent dynamics. 
Once quantized \cite{deWitt1967}, the solutions of this wave equation are spin-2 Bosons (gravitons), with dispersion  
\begin{eqnarray}
	\omega(\vb*{k}) = c |\vb*{k}| \; , 
\label{eqn.gr.dispersion}
\end{eqnarray}
and two independent polarizations, $\sigma=+,\times$, such that 
\begin{eqnarray}
   h_{\mu\nu}(t,\vb*{x}) &=& \sum_{\sigma=+,\times} \int d^3k  \frac{1}{\sqrt{\omega(\vb*{k})}} 
   \big[ \epsilon^\sigma_{\mu\nu} a^\dagger_\sigma (\vb*{k}) e^{ik_\rho x^\rho } \nonumber\\
   && + \quad \left( \epsilon^\sigma_{\mu\nu} \right)^* a^{\phantom\dagger}_\sigma (\vb*{k}) e^{-i k_\rho x^\rho} \big] \; ,
\label{eq:graviton}
\end{eqnarray} 
where $\epsilon^\sigma_{\mu\nu}$ is a tensor encoding information about polarization, and 
$a^{\phantom\dagger}_{\sigma}(\vb*{k})$ satisfies 
\begin{eqnarray}
	[ a^{\phantom\dagger}_{\sigma}(\vb*{k}), a^{\dagger}_{\sigma'}(\vb*{k'}) ] = \delta_{\sigma\sigma'} \delta (\vb*{k} - \vb*{k'}) \; .
\label{eq:bosonic.commutation.relation}
\end{eqnarray}  
For a wave with linear polarization, propagating along the $z-$direction,  
$\epsilon^\sigma_{\mu\nu}$ takes the specific form 
\begin{eqnarray}
\vb*{\epsilon}^+ = \frac{1}{\sqrt{2}} \begin{pmatrix}
0 & 0 & 0 & 0 \\
0 &1 & 0 & 0 \\
0 & 0 & -1 & 0 \\
0 & 0 & 0 & 0
\end{pmatrix},\ 
\vb*{\epsilon}^\times = \frac{1}{\sqrt{2}} \begin{pmatrix}
0 & 0 & 0 & 0 \\
0 & 0 & 1 & 0 \\
0 & 1 & 0 & 0 \\
0 & 0 & 0 & 0
\end{pmatrix} \; .
\label{eq:polarization.tensor}
\end{eqnarray}
%
%
Physically, this corresponds to a quadrupolar distortion of space, in which compression 
and dilation alternate.

\begin{figure}[t]
 	\begin{center}
 		\includegraphics[width=\columnwidth]{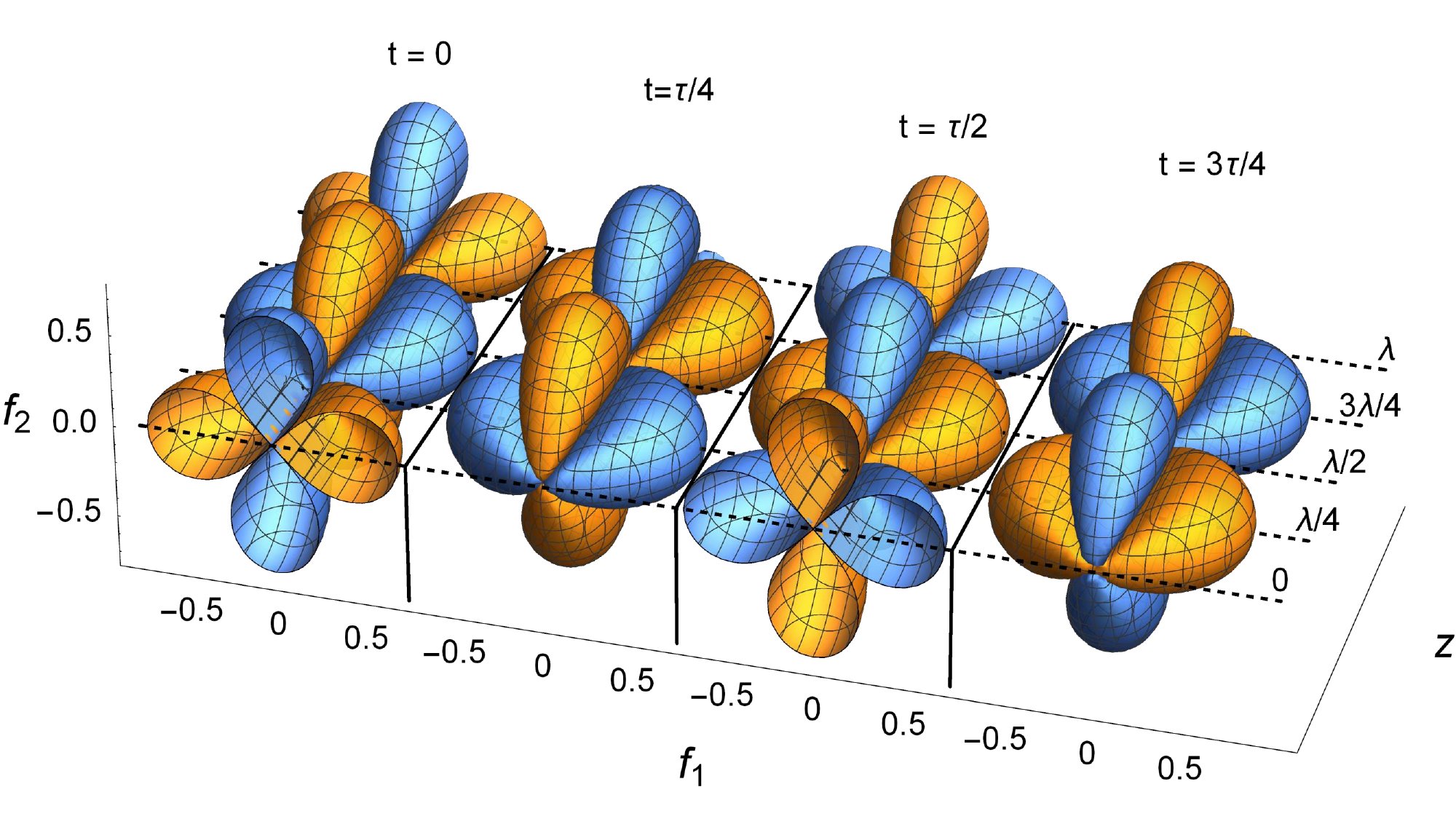}     
  		\centering
  		\caption{\small 
			Quadrupolar nature of gravitational waves, and Goldstone modes of spin-nematic order, 
			visualized through the associated distortions of spacetime, or the spin-nematic ground state.
			Results are shown for a wave of wavelength $\lambda$ and period 
			$\tau$, with polarization $\vb*{\epsilon}^+$ [Eq.~(\ref{eq:polarization.tensor})], 
			propagating along the z--axis. 
			In the case of gravitational waves [Eq.~(\ref{eq:graviton})], 
			$f_1, f_2$, represent $x$ and $y$ axes of spacetime, 
			and the quantity plotted is a surface of constant strain [Eq.~(\ref{eq:gr-strain},\ref{eq:gr-strain-surface})].
			In the case of spin--nematic order [Eq.~(\ref{eq:quadrupole.wave})], 
			$f_1, f_2$, represent spin components $S^x$ and $S^y$, 
			and the quantity plotted is the change in the spin--nematic order parameter 
			[Eq.~(\ref{eq:wavefunction.amplitude},\ref{eq:wavefunction-amplitude-surface})].
			Blue surfaces denote positive strain/deformation, while orange surfaces denote 
			negative strain/deformation.
  			An animated version of this figure is available in the Supplemental Materials~\cite{supplemental-material}. 
			}
 	\label{fig:local-excitations}
 	\end{center}
 \end{figure}

We define the strain (squared) to be 
\begin{align}
V(t,\vb*{x}) =   \frac{h_{mn}(t,\vb*{x}) x^m x^n}{|\vb*{x}|^2}.
\label{eq:gr-strain}
\end{align}
Here again, $m,n = 1,2,3$ run over the spatial components. 
In Fig.~\ref{fig:local-excitations} we visualize the equal strain surface in the $x^{1} - x^{2}$  plane (notation replaced by $f_1,\ f_2$), for a gravitational wave traveling in the $z-$direction, defined by 
\begin{equation}
V(t, f_1,f_2,z) = \pm \text{const.}
\label{eq:gr-strain-surface}
\end{equation} 

 
\textit{Linearized gravity analogue in spin nematics.--} 
%
In the discussion above, we have seen how small fluctuations of the metric $g_{\mu\nu}$, 
[Eq.~(\ref{eq:metric.tensor})], give rise to gravitational waves, which 
are linearly-dispersing massless spin-2 Bosons, described by the action $\mathcal{S}_\mathsf{LG}$ [Eq.~(\ref{eq:S.linearized.gravity})].
In this sense, the search for analogues of gravitational waves can be cast as the search for a physical system which can be described in terms of a symmetric, traceless \mbox{rank-2} tensor, with linearly-dispersing excitations 
governed by an action of the form $\mathcal{S}_\mathsf{LG}$.


Our strategy here is to map the gravitational perturbations on a flat background spacetime onto the Hilbert space of a quantum system whose ground state meets this description. 
%
Analogues of gravitational waves can then be found in the Goldstone modes of this 
symmetry-broken state.


Classically, the order parameter for a nematic liquid crystal 
is a symmetric, traceless \mbox{rank-2} tensor \cite{Frank1958}. 
We here consider a form of quantum liquid crystal known as a ``quantum spin nematic'', originally introduced 
as a magnetic state \cite{Andreev1984,Chubukov1991,Shannon2006} which preserves time-reversal 
symmetry, but breaks spin-rotation symmetry through the quadrupole operators 
\begin{align}
{\mathcal Q}^{mn} 
= \frac{1}{2} \left( S^m S^n 
+ S^n S^m \right) 
- \frac{1}{3} \delta_{mn} S^n S^n \; .
\label{eq:nematic.order.parameter}
\end{align} 
Here $S^m$ is a spin operator with components $m=x,y,z$, 
satisfying the usual SU(2) commutation relations. 

\begin{figure}[t]
	\centering
	\subfloat[ FQ ground state \label{fig:FQ.state} ]{\includegraphics[width=0.45\columnwidth]{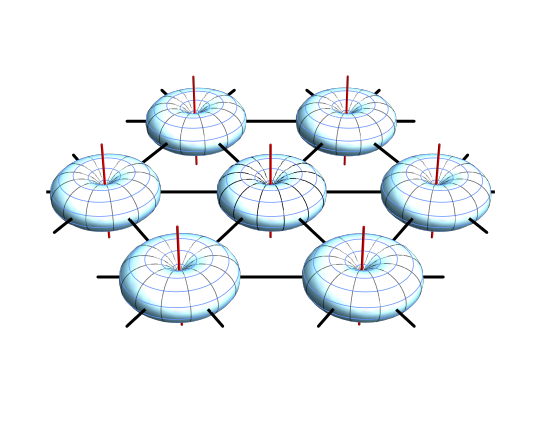}}
	\subfloat[ $S_{\sf Q} (\vb*{k}, \omega)$ \label{fig:FQ.structure.factor} ]{\includegraphics[width=0.65\columnwidth]{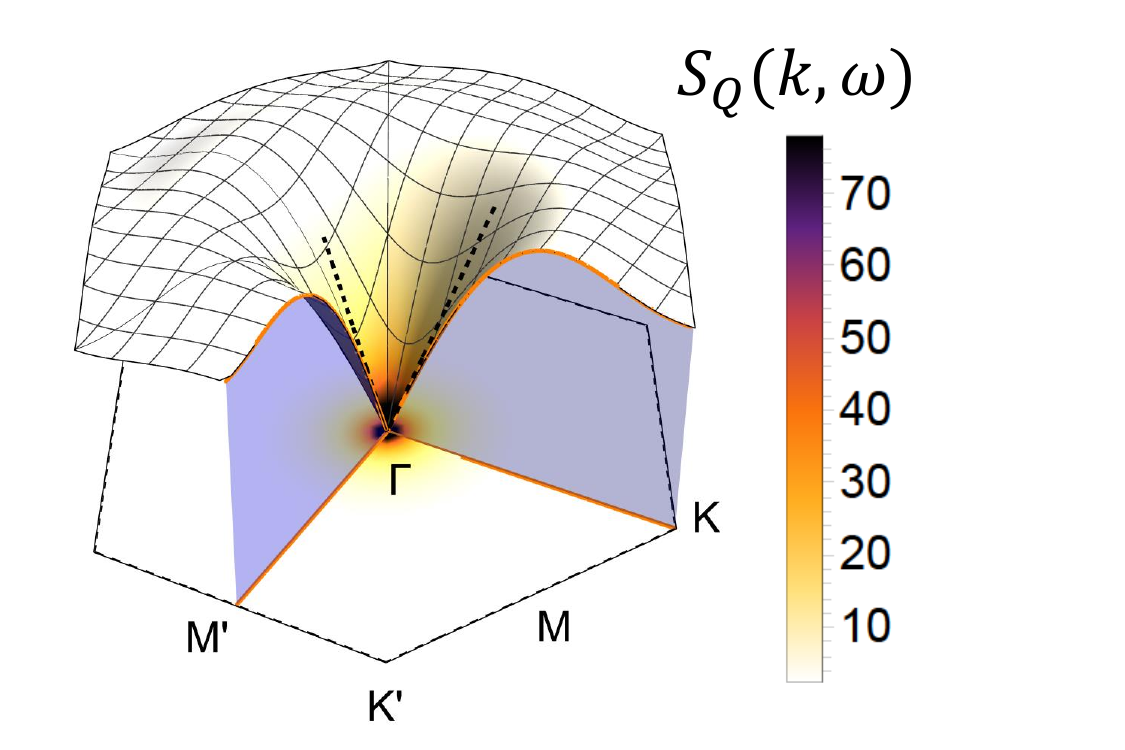}}
	\caption{\small  
		Spin--nematic state on a triangular lattice, and its \mbox{spin-2} excitations.   
		(a) Ferroquadrupolar (FQ) ground state, in which quadrupole moments of spin align with a common axis.
		(b)  Dispersion of excitations about the FQ state, showing linear character $\omega = v |\vb*{k}|$ 
		at long wavelength (black dashed line).
		This linear dispersion is consistent with the predictions of the field theory [Eq.~(\ref{eq:S.FQ})].
		The spin--2 nature   of the long-wavelength excitations can be inferred from the quadrupolar structure factor 
		$S_{\rm Q}(\vb*{k}, \omega)$ [Eq.~(\ref{eqn.sqw})], overlaid on the plot.
		Results are shown for a spin-1 bilinear biquadratic (BBQ) model  [Eq.~(\ref{eq:H.BBQ})], 
		with parameters $J_1=0, J_2 = -1$, as described in \cite{Remund2022}.
	}
\end{figure}

The simplest form of a quantum spin nematic is the ``ferroquadrupolar'' (FQ) state, a uniaxial nematic 
liquid crystal in which all quadrupole moments are aligned [Fig.~\ref{fig:FQ.state}].
As in conventional liquid crystals \cite{Mermin1979}, such a state can be characterized by a director ${\vb* d}$, 
and its symmetry dictates that it supports two, degenerate Goldstone modes \cite{watanabe_unified_2012}, 
which have the character of massless, spin-2 Bosons \cite{Ivanov2003,Lauchli2006,Remund2022}. 
We will now show how these correspond to the massless spin--2 Bosons found in linearised gravity.


We start by promoting ${\mathcal Q}^{mn}$ to a tensor field $Q^{\mu\nu}$ providing a low energy effective description, identifying \mbox{$Q^{mn} = {\mathcal Q}^{mn}$}, where \mbox{$Q^{mn} = Q_{mn}$}, and by setting components \mbox{$Q_{0\mu} = Q_{\mu0} = 0$}.
%
%
In analogy with Eq.~(\ref{eq:metric.tensor}), we consider fluctuations $Q^{\sf E}_{\mu\nu}$
about a state with uniform spin nematic order ${Q}^{\sf GS}_{\mu\nu}$, viz 
\begin{eqnarray}
	Q_{\mu\nu} =  Q^{\sf GS}_{\mu\nu} + Q^{\sf E}_{\mu\nu} \; ,
\label{eq:decomposition.of.Q}
\end{eqnarray}
requiring that these fluctuations occur in the transverse channel, 
i.e., that the change affects the direction but not the magnitude of the quadrupolar order.
This assumption is appropriate for the low-energy physics of spin nematics 
\cite{Ivanov2003,Smerald2013,Smerald2013-book}.
 

What remains is to match the fluctuations of quadrupolar order, which occur in spin--space, to the 
changes in spacetime coordinates appropriate for a gravitational wave.   
This can be accomplished by a unitary transformation 
\begin{align}
	\tilde{Q}_{\mu\nu}(\vb*{k})  = C_{\mu\nu}{}^{\rho\sigma}(\vb*{k},\vb*{d}) Q^{\sf E}_{\rho\sigma}(\vb*{d}) \; ,
\end{align}
where $C_{\mu\nu}{}^{\rho\sigma}(\vb*{k},\vb*{d})$ acts on quadrupole excitations with wave vector 
$\vb*{k}$ about a FQ state characterized by director $\vb*{d}$.
Further details of this transformation are given in the Supplemental Material \cite{supplemental-material}.


For an appropriate choice of $C_{\mu\nu}{}^{\rho\sigma}(\vb*{k},\vb*{d})$, $\tilde{Q}_{\mu\nu}$  satisfies the conditions
\begin{subequations}
	\begin{align}	
	&\tilde{Q}^\mu_{\ \mu}  = Q^\mu_{\ \mu}     = 0 \; \textrm{[traceless]}, \\
	& \tilde{Q}_{0\mu}  = Q_{0\mu} = 0  \;  \textrm{[no scalar or vector components]}, \\
	&k_m \tilde{Q}_{mn}(k^\sigma) = 0 \; \textrm{[no longitudinal dynamics]}.
	\label{eq:constraints-FQ}
	\end{align}
\end{subequations}


The low-energy fluctuations of the spin nematic can be described
in terms of a quantum non-linear sigma model \cite{Ivanov2003,Smerald2013,Smerald2013-book}.   
Given these physical constraints, and the decomposition described by Eq.~(\ref{eq:decomposition.of.Q}), 
we arrive at an action which exactly parallels linearized gravity [Eq.~(\ref{eq:S.linearized.gravity})], via, 
\begin{eqnarray}
\begin{split}
  \mathcal{S}_\mathsf{FQ} 
    = -\frac{1}{2}\int dtd^d x \bigg[ 
    & \chi_{\perp}(\partial^t \tilde{Q}^{\mu\nu}\partial_t \tilde{Q}_{\mu\nu}) \\ 
    & - \rho_{s}(\partial^n \tilde{Q}^{\mu\nu}\partial_n \tilde{Q}_{\mu\nu})\bigg],
\end{split}
\label{eq:S.FQ}
\end{eqnarray}
where $\chi_{\perp}$ is the transverse susceptibility, and $\rho_{s}$ the stiffness, associated 
with spin-nematic order \cite{Smerald2013,Smerald2013-book}.


The low-lying excitations of this theory are massless spin-2 Bosons, satisfying the wave 
equation [cf. Eq.~(\ref{eq:EoM.linearized.gravity})]
%
\begin{eqnarray}
	\frac{1}{v^2}\partial_t\partial^t \tilde{Q}_{\mu\nu} &-& \partial_n\partial^n \tilde{Q}_{\mu\nu} = 0 \; , 
	\label{eq:EoM.spin.nematic} 
\end{eqnarray}
%
with $v = \sqrt{{\rho_{s}}/{\chi_{\perp}}} $, and with dispersion
\begin{align}
	\omega (\vb*{k}) = v |\vb*{k}|  \; .
\label{eqn.fq.dispersion}
\end{align}
The solutions to Eq.~(\ref{eq:EoM.spin.nematic}) 
have exactly the same structure as those for gravitons [cf. Eq.~(\ref{eq:graviton})]
\begin{equation}
\begin{split}
\tilde{Q}_{\mu\nu}(t,\vb*{x}) 
	= \sum_{\sigma=+,\times} \int d^3k \frac{1}{\sqrt{\omega (\vb*{k}) }} 
	\big[ & \epsilon^\sigma_{\mu\nu} b^\dagger_\sigma (\vb*{k}) e^{ik_\rho x^\rho } \\
	& + \left( \epsilon^\sigma_{\mu\nu} \right)^* b^{\phantom\dagger}_\sigma (\vb*{k}) e^{-ik_\rho x^\rho } \big] \; ,
\end{split}
\label{eq:quadrupole.wave}
\end{equation}
where $b^{\phantom\dagger}_\sigma(\vb*{k})$ satisfies Bosonic commutation 
relations [Eq.~(\ref{eq:bosonic.commutation.relation})] and,  
the tensors $\vb*{\epsilon}^{\sigma}_{\mu\nu}$ are given by Eq.~(\ref{eq:polarization.tensor}).


The quadrupolar excitations can be visualized in terms of surfaces proportional to the wavefunction amplitudes of the propagating mode, in analogy with Eq.~(\ref{eq:gr-strain})
\begin{align}
	V ( \vb*{S}, (t, \vb*{x}))= \frac{S^m \tilde{Q}_{mn}(t,\vb*{x})S^n}{|\vb*{S}|^2}.
\label{eq:wavefunction.amplitude}
\end{align} 
In Fig.~\ref{fig:local-excitations} we plot the equal amplitude surface in the $S^x-S^y$ plane (notation changed to $f_1,\ f_2$), for a quadrupole wave propagating in the $z-$direction, defined as 
\[
	V ( \vb*{S} = (f_1, f_2,0), (t,0,0,z))= \pm \text{const.}
\label{eq:wavefunction-amplitude-surface}
\]
as shown in Fig.~\ref{fig:local-excitations}.


It follows that, from a mathematical point of view, the quadrupolar waves in a quantum 
spin nematic are in one-to-one correspondence with quantized gravitational waves (gravitons) 
in a flat, 4-dimensional spacetime.
However, there is a critical distinction regarding the spaces these waves arise in, which has important 
implications for realizing them in experiment.
Gravitational waves involve quadrupolar distortions of space, transverse to the direction of propagation.
This implies that a minimum of three spatial dimensions is required to support a gravitational wave.
In contrast, the quadrupolar waves found in spin nematics arise in a spin--space which is automatically 
three--dimensional, regardless of the number of spatial dimensions.
For this reason, it is possible to explore analogues of gravitational waves in 2-dimensional spin systems.  
It is this subject which we turn to next.


\begin{figure*}[th!]
	\begin{center}
		\includegraphics[width=\textwidth]{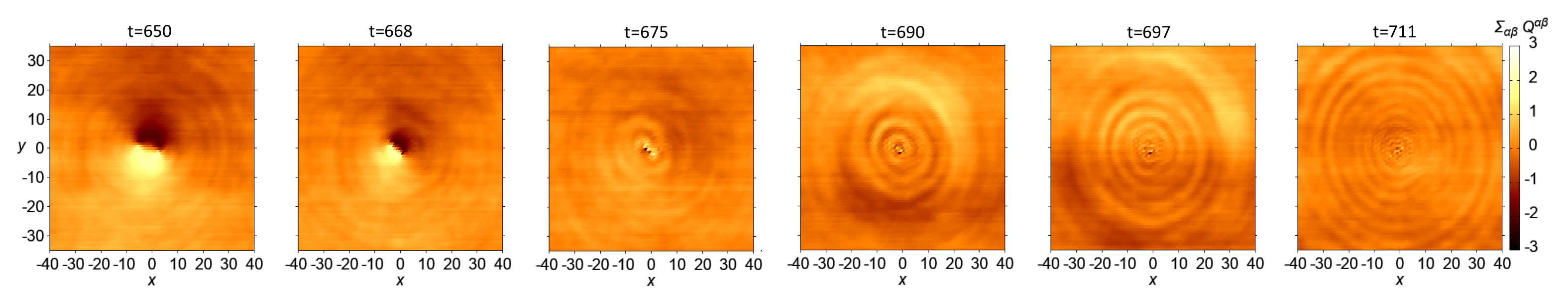}  
		\caption{\small 
		Numerical simulation of vortices within a spin-nematic state, showing how quadrupole waves, 
		analogous to gravitational waves, are created when a pair of vortices in-spiral and annihilate.
		Individual frames are taken from dynamical simulation of a ferroquadrupolar state (FQ) in the 
		spin-1 bilinear biquadratic (BBQ) model on a triangular lattice [Eq.~(\ref{eq:H.BBQ})], 
		with further detail
		given in the Supplemental Material \cite{supplemental-material}.
		An animated version of this result is also available \cite{supplemental-material}.
		}
    \label{fig:simulation-result}
    \end{center}
\end{figure*}

\textit{Simulation using cold atoms.-- }
%
The idea of using cold atoms to simulate a quantum spin nematic has a long history
\cite{Ohmi1998,Demler2002,Imambekov2003,deParny2014,Zibold2016}.
The majority of proposals build on ``spinor condensates'' of atoms, such as $^{23}$Na, $^{39}$K or $^{87}$Rb, 
whose internal hyperfine states mimic the magnetic basis of a spin-1 moment \cite{Ohmi1998,Ho1998,Stenger1998}.
The interactions between these effective \mbox{spin-1} moments depend on the details of their scattering 
and, where attractive, can lead to spin-nematic order \cite{Ohmi1998}.
Condensates described by the order parameter ${\mathcal Q}^{\alpha\beta}$ [Eq.~(\ref{eq:nematic.order.parameter})] 
have already been observed in experiment \cite{Jacob2012}.
On symmetry grounds, the Goldstone modes of these systems must be described by $\mathcal{S}_\mathsf{FQ}$ [Eq.~(\ref{eq:S.FQ})], 
making them analogues of linearized gravity.


Optical lattices can be arranged in a wide array of geometries, including triangular lattices \cite{Becker2010}, 
and cold atom experiments with $^{23}$Na atoms are carried out in many laboratories, e.g., \cite{Xu2005,Zibold2016,fava_observation_2018}.  
Realizing an analogue of gravitational waves on a lattice therefore also seems a realistic possibility. 


\textit{Realization of gravitational waves in a microscopic lattice model.-- } 
In addition to realization of analogue gravitational waves using quantum fluids as suggested above, 
spin-nematic phases can also be found in solid state magnetic systems. 
The simplest microscopic model supporting a quantum spin-nematic state is the spin-1 bilinear 
biquadratic (BBQ) model 
\begin{align}
	{\mathcal H}_\mathsf{BBQ} 
	= J_1 \sum_{\langle ij \rangle} \vb*{S}_i \cdot \vb*{S}_j 
	+ J_2 \sum_{\langle ij \rangle} \big( \vb*{S}_i \cdot \vb*{S}_j \big)^2  \; ,
\label{eq:H.BBQ}
\end{align}
known to support FQ order for a wide range of $J_2 < 0$, 
irrespective of lattice geometry \cite{Andreev1984,Papanicolaou1988,Harada2002}.
Particular attention has been paid to the BBQ model on a triangular lattice \cite{Lauchli2006,Tsunetsugu2006,Stoudenmire2009,Kaul2012,Smerald2013,Voll2015,Remund2022}, 
where studies have been motivated by e.g. 
NiGa$_2$S$_4$ 
\cite{nakatsuji_spin_2005,valentine_impact_2020}, 
and FeI$_2$ \cite{bai_hybridized_2021}. 
It has also been argued that $^{23}$Na atoms in an optical lattice could be used to realize the 
BBQ model [Eq.~(\ref{eq:H.BBQ})], with parameters falling into the range relevant to FQ 
order \cite{Imambekov2003,StamperKurn2013}.  


Explicit calculations of FQ dynamics within the BBQ model reveal two, degenerate 
bands of excitations, with linear dispersion at long wavelength \cite{Lauchli2006,Voll2015,Remund2022}.  
The quadrupolar (spin-2) nature of these excitations at low energy
is manifest in the dynamical structure factor for quadrupole moments
\begin{equation}
	\label{eqn.sqw}
    S_{\mathcal Q}(\vb*{k}, \omega) = \sum_{\alpha,\beta} \int \frac{dt}{2\pi} e^{i \omega t} \langle {\mathcal Q}^{\alpha\beta}(\vb*{k}, t) {\mathcal Q}^{\alpha\beta}( \vb*{-k},0)  \rangle \; ,
 \end{equation}
shown in Fig.~\ref{fig:FQ.structure.factor}, for calculations carried out at a semiclassical 
level \cite{Remund2022}.   
Starting from Eq.~(\ref{eq:H.BBQ}), it is also possible to parameterize the continuum 
field theory Eq.~(\ref{eq:S.FQ}), obtaining results in quantitative agreement with the microscopic model, as shown 
in Fig.~\ref{fig:FQ.structure.factor}.

\textit{Quench dynamics, simulation and measurement.-- }
%
We now turn to the question of how gravitational-wave analogues could be created and observed in experiment.  
For concreteness, we consider a FQ state in an explicitly 2-dimensional system, which we model as set of
spin-1 moments on a lattice [cf.~Eq.~(\ref{eq:H.BBQ})].
Consistent with the Mermin-Wagner theorem, for low-dimensional systems to exhibit anything besides exponentially-decaying correlations at low temperature, they must undergo topological phase transitions e.g. of the BKT type \cite{Berezinski1972,Kosterlitz1973}. 
The FQ state in a 2D magnet is known to be connected to the high temperature magnetic phase via a vortex-induced topological phase transition \cite{Kawamura2007}.


The excitations which mediate this phase transition are no longer the integer vortices of the Berezinskii-Kosterlitz-Thouless (BKT) transition \cite{Hadzibabic2006}, but rather $\mathcal{Z}_2$ vortices of homotopy group $\pi_1$, specific to the nematic order parameter \cite{Mermin1979}.
Cooling rapidly through the transition (quenching) leads to a state 
rich in pairs of $\mathcal{Z}_2$ vortices, which are subject to attractive interactions, and spiral towards 
one another in much the same way as gravitating masses.
In the process, vortices radiate energy in the form of quadrupolar waves, Eq.~(\ref{eq:quadrupole.wave}), and eventually annihilate.
This process is clearly visible in simulations of the BBQ model [Eq.~(\ref{eq:H.BBQ})], as illustrated 
in Fig.~\ref{fig:simulation-result} \cite{supplemental-material}, and the accompanying 
animation \cite{supplemental-material}.


As can be seen from these simulations, the dynamics of vortices is very slow compared to that of quadrupolar 
waves, and the timescale associated with the annihilation of $\mathcal{Z}_2$ vortices is of order $10^2 J_2^{-1}$. 
Observing vortices in experiment will therefore typically demand long--lived condensates.
None the less, successful imaging of conventional magnetic vortices within a spinor--condensate 
of $^{23}$Na ions has already been realized, over timescales of \text{$\sim 1\text{s}$} \cite{Kang2019PhysRevLett}.
An experimental protocol for observing excitations in the quadrupolar channel of a spinor condensate 
has also already been implemented \cite{Kunkel2019}, and could be used to make real--space images 
of nematic correlations, through their imprint on the electric polarisability of the atoms \cite{Carusotto2004}.
Proposals also exist for imaging the quadrupolar correlations of spin-1 moments through, e.g., 
Raman scattering \cite{Michaud2011}.  

Taken together, this all jointly suggests that it is a realistic possibility to realize and observe such spin nematic gravitational waves analogues. 

%
\textit{Conclusions. -- }
Linearized gravity is of fundamental interest, but hard to study in experiment, because the 
energy scales of excitations are so large, and their amplitudes so small.
In this Letter we have shown how a theory in direct correspondence to linearized 
gravity arises in systems with spin-nematic order.
The Goldstone modes of this spin-nematic state are massless spin-2 Bosons, which 
behave as exact analogues of quantized gravitational waves (gravitons).   
These results imply that it is possible to simulate various aspects of linearized gravity, 
including gravitons and topological excitations, in magnets or assemblies of cold atoms which realize a spin-nematic state \cite{IvanovPhysRevLett.100.047203,UedaPhysRevA.93.021606,AkagiPhysRevD.103.065008,Akagi2021JHEP,AmariPhysRevB.106.L100406}.

\begin{acknowledgments}
\textit{Acknowledgments -- }
The authors are pleased to acknowledge helpful discussions with Yuki Amari, Andrew Smerald and Hiroaki T. Ueda.
This work was supported by Japan Society for the Promotion of Science (JSPS) KAKENHI Grants No. 
JP19H05822,  
JP19H05825,  
JP20K14411,  
JP20H05154 and 
JP22H04469, 
MEXT "Program for Promoting Researches on the Supercomputer Fugaku", Grant No. JPMXP1020230411, 
JST PRESTO Grant No. JPMJPR2251,  
and by the Theory of Quantum Matter Unit, OIST. 
Numerical calculations were carried out using HPC facilities provided by OIST 
and the Supercomputer Center of the Institute of Solid State Physics, the University of Tokyo.
%
\end{acknowledgments}
 
\bibliography{paper.bib}

\clearpage


\section*{Supplementary material}

# Supplemental material: Gravitational wave analogues in spin nematics and cold atoms

Leilee Chojnacki, Rico Pohle, Han Yan, Yutaka Akagi, and Nic Shannon
(Dated: October 16, 2023)

## I. DICTIONARY OF THE TRANSFORMATION FROM SPIN–SPACE TO SPACETIME

Ferroquadrupolar (FQ) order is a broken–symmetry state in which quadrupolar degrees of freedom align to select a unique axis [1, 2]. This axis is characterized by a director

$$\boldsymbol{d} = (d_1, d_2, d_3) \, , \tag{S1}$$

which we can choose to be of unit magnitude. Then, for a FQ ground state and selecting the $z$–axis, we have

$$\boldsymbol{d} = (0, 0, 1) = \hat{\boldsymbol{z}} \, , \tag{S2}$$

and the tensor describing quadrupolar order [Eq. (12,13) of main text] is

$$Q^{\mathsf{GS}}_{\mu\nu} = \begin{pmatrix} 0 & 0 & 0 & 0 \\ 0 & -\frac{1}{3} & 0 & 0 \\ 0 & 0 & -\frac{1}{3} & 0 \\ 0 & 0 & 0 & \frac{2}{3} \end{pmatrix} . \tag{S3}$$

Since FQ order breaks spin–rotation symmetry, it must possess gapless Goldstone modes. These involve fluctuations of the quadrupole moment perpendicular to the director defining the ground state. In the present example, for $\boldsymbol{d} \parallel \hat{\boldsymbol{z}}$, the relevant components of the tensor are $Q_{xz} = Q_{zx}$ and $Q_{yz} = Q_{zy}$, leading to modes $Q^{\mathsf{E}}_{\mu\nu}$ with character

$$\boldsymbol{\epsilon}^{xz} \sim \begin{pmatrix} 0 & 0 & 0 & 0 \\ 0 & 0 & 0 & 1 \\ 0 & 0 & 0 & 0 \\ 0 & 1 & 0 & 0 \end{pmatrix}, \qquad \boldsymbol{\epsilon}^{yz} \sim \begin{pmatrix} 0 & 0 & 0 & 0 \\ 0 & 0 & 0 & 0 \\ 0 & 0 & 0 & 1 \\ 0 & 0 & 1 & 0 \end{pmatrix}, \tag{S4}$$

(c.f. Section IVA in [2]). Since quadrupole moments are defined in spin–space, not spacetime, the form of this excitation does not depend on the direction that the wave is propagating along.

This contrasts with the description of a gravitational wave given in Eq. (7) of the main text. For the specific case of a wave propagating along the $z$–axis, we have a wave vector

$$\boldsymbol{k} = (0, 0, k) = k\hat{\boldsymbol{z}} \, , \tag{S5}$$

and the associated, quadrupolar, fluctuations of spacetime involve different components of $h_{\mu\nu}$, viz

$$\boldsymbol{\epsilon}^{+} \sim \begin{pmatrix} 0 & 0 & 0 & 0 \\ 0 & 1 & 0 & 0 \\ 0 & 0 & -1 & 0 \\ 0 & 0 & 0 & 0 \end{pmatrix}, \qquad \boldsymbol{\epsilon}^{\times} \sim \begin{pmatrix} 0 & 0 & 0 & 0 \\ 0 & 0 & 1 & 0 \\ 0 & 1 & 0 & 0 \\ 0 & 0 & 0 & 0 \end{pmatrix} . \tag{S6}$$

We note that, for gravitational waves, the structure of the excitation *does* depend on the direction of propagation, since the wave itself is a fluctuation of spacetime.

Superficially, these might appear to be different excitations. However, these differences stem from the different coordinate systems used to describe excitations in spacetime, and excitations within the order-parameter space associated with the spin-quadrupole moment. When we talk about equivalence between gravitational waves and quadrupole waves, we are using the internal, spin degrees of freedom at a given point in space and time as a "notebook" to store information about the linearized fluctuations of spacetime. We are free to transcribe these fluctuations of spacetime as we think fit, as long as we preserve their structure as massless spin–2 Bosons. And this means we have the freedom to chose a coordinate system in which the correspondence between the Goldstone mode and the gravitational wave becomes apparent.

These different, but physically equivalent, representations are linked by a unitary transformation acting on tensors in spin space. For the example considered above, with $\boldsymbol{k} \parallel \boldsymbol{d} \parallel \hat{\boldsymbol{z}}$, we write

$$\tilde{Q}_{\mu\nu}(k\hat{\boldsymbol{z}}) = C_{\mu\nu}{}^{\rho\sigma}(\boldsymbol{k} = k\hat{\boldsymbol{z}}, \boldsymbol{d} = \hat{\boldsymbol{z}}) \, Q^{E}_{\rho\sigma}(\hat{\boldsymbol{z}}) \, , \tag{S7}$$

where $C_{\mu\nu}{}^{\rho\sigma}(\boldsymbol{k} = k\hat{\boldsymbol{z}}, \boldsymbol{d} = \hat{\boldsymbol{z}})$ is a rank-4 tensor given by

$$C_{\mu\nu}{}^{\rho\sigma}(\boldsymbol{k} = k\hat{\boldsymbol{z}}, \boldsymbol{d} = \hat{\boldsymbol{z}}) = [\lambda_1]_{\mu\nu}[\lambda_4]^{\rho\sigma} + [\lambda_3^{xy}]_{\mu\nu}[\lambda_6]^{\rho\sigma} = \lambda_1 \otimes \lambda_4 + \lambda_3^{xy} \otimes \lambda_6 \;, \tag{S8}$$

and $\lambda_i$ are a subset of generators of $SU(3)$, padded with zeros, viz

$$\lambda_1 = \begin{pmatrix} 0 & 0 & 0 & 0 \\ 0 & 0 & 1 & 0 \\ 0 & 1 & 0 & 0 \\ 0 & 0 & 0 & 0 \end{pmatrix}, \quad \lambda_6 = \begin{pmatrix} 0 & 0 & 0 & 0 \\ 0 & 0 & 0 & 0 \\ 0 & 0 & 0 & 1 \\ 0 & 0 & 1 & 0 \end{pmatrix}, \quad \lambda_4 = \begin{pmatrix} 0 & 0 & 0 & 0 \\ 0 & 0 & 0 & 1 \\ 0 & 0 & 0 & 0 \\ 0 & 1 & 0 & 0 \end{pmatrix},$$

$$\lambda_3^{xy} = \begin{pmatrix} 0 & 0 & 0 & 0 \\ 0 & 1 & 0 & 0 \\ 0 & 0 & -1 & 0 \\ 0 & 0 & 0 & 0 \end{pmatrix}, \quad \lambda_3^{yz} = \begin{pmatrix} 0 & 0 & 0 & 0 \\ 0 & 0 & 0 & 0 \\ 0 & 0 & 1 & 0 \\ 0 & 0 & 0 & -1 \end{pmatrix}, \quad \lambda_3^{xz} = \begin{pmatrix} 0 & 0 & 0 & 0 \\ 0 & 1 & 0 & 0 \\ 0 & 0 & 0 & 0 \\ 0 & 0 & 0 & -1 \end{pmatrix}. \tag{S9}$$

Being real-valued, these generators have the effect of rotating quadrupole moments, without introducing a dipole moment [3]. Crucially, they preserve both the magnitude of the quadrupole moment, and the structure of the order–parameter manifold, $RP(2)$, allowing us to use the same non-linear sigma model description [Eq. (16) of main text] for both $Q^{\sf E}_{\rho\sigma}$ and $\tilde{Q}_{\mu\nu}$.

We can now generalize to the case of a gravitational wave propagating in an arbitrary direction $\hat{\boldsymbol{k}}$, as referred to FQ order with arbitrary orientation of director $\hat{\boldsymbol{d}}$. This can be accomplished by rotating the spatial components of the coordinate system so as to align $\hat{\boldsymbol{k}}$ and $\hat{\boldsymbol{d}}$ with a common $z$–axis. For the general case [Eq. (14) of main text] we write

$$\tilde{Q}_{\mu\nu}(\boldsymbol{k}) = C_{\mu\nu}{}^{\rho\sigma}(\boldsymbol{k}, \boldsymbol{d}) \, Q^E_{\rho\sigma}(\boldsymbol{d}) \;, \tag{S10}$$

where the transformation $C_{\mu\nu}{}^{\rho\sigma}(\boldsymbol{k}, \boldsymbol{d})$ is given by

$$C_{\mu\nu}{}^{\rho\sigma}(\boldsymbol{k}, \boldsymbol{d}) = R[\boldsymbol{m}, \phi]^{\alpha}{}_{\mu} R[\boldsymbol{m}, \phi]^{\beta}{}_{\nu} R[\boldsymbol{n}, \theta]^{\gamma}{}_{\rho} R[\boldsymbol{n}, \theta]^{\delta}{}_{\sigma} \left[ \lambda_1 \otimes \lambda_4 + \lambda_{3xy} \otimes \lambda_6 \right]_{\alpha\beta\gamma\delta} \;, \tag{S11}$$

with

$$\boldsymbol{m} = \frac{\boldsymbol{k} \times \hat{\boldsymbol{z}}}{|\boldsymbol{k} \times \hat{\boldsymbol{z}}|} \;, \qquad \cos\phi = \hat{\boldsymbol{k}} \cdot \hat{\boldsymbol{z}} \;, \quad \boldsymbol{n} = \frac{\boldsymbol{d} \times \hat{\boldsymbol{z}}}{|\boldsymbol{d} \times \hat{\boldsymbol{z}}|} \;, \qquad \cos\theta = \hat{\boldsymbol{d}} \cdot \hat{\boldsymbol{z}} \;, \tag{S12}$$

and the rotation matrix

$$\boldsymbol{R}[\boldsymbol{m}, \theta] = \begin{pmatrix} 0 & 0 & 0 & 0 \\ 0 & \mathrm{Cos}[\theta] + (1 - \mathrm{Cos}[\theta])m_1^2 & (1 - \mathrm{Cos}[\theta])m_1 m_2 - \mathrm{Sin}[\theta]m_3 & \mathrm{Sin}[\theta]m_2 + (1 - \mathrm{Cos}[\theta])m_1 m_3 \\ 0 & (1 - \mathrm{Cos}[\theta])m_1 m_2 + \mathrm{Sin}[\theta]m_3 & \mathrm{Cos}[\theta] + (1 - \mathrm{Cos}[\theta])m_2^2 & -\mathrm{Sin}[\theta]m_1 + (1 - \mathrm{Cos}[\theta])m_2 m_3 \\ 0 & -\mathrm{Sin}[\theta]m_2 + (1 - \mathrm{Cos}[\theta])m_1 m_3 & \mathrm{Sin}[\theta]m_1 + (1 - \mathrm{Cos}[\theta])m_2 m_3 & \mathrm{Cos}[\theta] + (1 - \mathrm{Cos}[\theta])m_3^2 \end{pmatrix}. \tag{S13}$$

Defined in this way, a separate transformation needs to be carried out for each wavevector $\boldsymbol{k}$. However, because the sigma model $\mathcal{S}_{\sf FQ}$ [Eq. (16) of main text] is invariant under both rotations of space, and rotations within the ground state manifold, these transformations do not alter our description of the Goldstone modes of FQ order.

One may note that the rank-4 transformation matrix $C_{\mu\nu}{}^{\rho\sigma}(\boldsymbol{k}, \boldsymbol{d})$ is $\boldsymbol{k}$–dependent and ill-defined at the limit $\boldsymbol{k} \to \boldsymbol{0}$. However, this does not pose a problem. After all, the gravitational wave gauge fixing $k^n h_{mn} = 0$ (Eq. (3c) in the main text) is also singular at $\boldsymbol{k} \to \boldsymbol{0}$ and ill-defined. So we cannot expect the nematic quadrupole wave to work there neither. Away from $\boldsymbol{k} = \boldsymbol{0}$, the system decouples into independent harmonic oscillators at each $\boldsymbol{k}$, and the transformation matrix $C_{\mu\nu}{}^{\rho\sigma}(\boldsymbol{k}, \boldsymbol{d})$ varies smoothly while mapping one harmonic oscillator to the other. The identification of the two theories hence does not suffer any ill-definition, discontinuity, etc. The only exception is when $\boldsymbol{k} \propto -\hat{\boldsymbol{z}}$. This will be discussed in detail in a future work.

## II. VISUALIZATION OF GRAVITATIONAL WAVES AND EQUIVALENT GOLDSTONE MODES

Linearized gravity [Eq. (4) of main text] supports gravitational waves, which take the form of fluctuations of the metric tensor $h_{\mu\nu}$ [Eq. (7) of main text]. Meanwhile, FQ order [Eq. (12, 13) of main text] supports Goldstone modes which are fluctuations of the order parameter $\tilde{Q}_{\mu\nu}$ [Eq. (19) of main text]. In the First Animation (Supplemental Materials), and Fig. 1 of the main text, we visualize these excitations through:

1. The strain caused by the passage of a gravitational wave. In this case the axes $f_1 = x$ and $f_2 = y$ refer to real–space, and the wave is visualized through a surface of constant strain [Eq. (10, 11) of the main text].

2. The amplitude of fluctuations about FQ order. In this case the axes $f_1 = S_x$ and $f_2 = S_y$ refer to spin–space, and the wave is visualized through a surface of constant amplitude [Eq. (20,21) of the main text].

In both cases, we consider a wave of wavelength $\lambda$ and period $\tau = \frac{2\pi}{\omega} = \frac{\lambda}{c}$, with polarization $\epsilon^+$ [Eq. (9) of main text]. The relative phase of fluctuations is shown using color, with blue surfaces denoting positive strain/amplitude and orange surfaces denoting negative strain/amplitude. Both the gravitational wave, and the FQ Goldstone mode, exhibit identical quadrupolar fluctuations, within their respective spaces.

### III. NUMERICAL SIMULATION OF VORTICES WITHIN FQ STATE

The numerical simulations shown in the Second Animation (Supplemental Materials) and Fig. 3 of the main text, were carried out using (semi–)classical methods developed and documented in [2]. The model considered was the spin-1 bilinear–biquadratic (BBQ) model on a triangular lattice [Eq. (21) of the main text], with parameters $J_1 = 0$, $J_2 = -1$. For these parameters, the BBQ model has a FQ ground state [1], which supports both quadrupolar Goldstone modes [1, 2], and point–like $Z_2$ vortices [4].

In order to exhibit the dynamics of vortices, an initial state was prepared containing two vortices

$$\boldsymbol{d}(\boldsymbol{r}) = \begin{cases} (-\sin\theta(\boldsymbol{r}), \cos\theta(\boldsymbol{r}), 0) & \text{for } x_1 \leq x \leq x_2 \\ (\cos\theta(\boldsymbol{r}), \sin\theta(\boldsymbol{r}), 0) & \text{otherwise} \end{cases}, \quad (S14)$$

where, by analogy with defects in classical 2D nematic liquid crystals [5], we write

$$\theta(\boldsymbol{r}) = \frac{1}{2}\tan^{-1}\left(\frac{y - y_1}{x - x_1}\right) - \frac{1}{2}\tan^{-1}\left(\frac{y - y_2}{x - x_2}\right). \quad (S15)$$

Here $\boldsymbol{r} = (x, y)$, and vortices are located at $\boldsymbol{r}_1$, $\boldsymbol{r}_2$. This ansatz, illustrated in Fig. S1, describes vortices with core of zero radius (point defects), and common orientation.

Starting from this initial state, numerical integration of equations of motion was then performed using an $4^{th}$ order Runge-Kutta (RK4) algorithm, with a total of $N_t = 7000$ time steps, each of duration $\delta t = 0.05 J_2^{-1}$. This gives a total simulation time of $\Delta t = 350 J_2^{-1}$. For purposes of animation, this time-series was sampled once every 7 time steps, and frames labeled $t = 1, \ldots, 1000$ [cf. Fig. 3 of the main text]. Vortices are subject to an attractive interaction, and over the course of simulations, they are seen to spiral towards one another, eventually annihilating. Over the course of this process, energy is radiated from the vortex pair in the form of quadrupolar waves.

We have also carried out simulations of the FQ phase of the BBQ model at finite temperature, using the classical Monte Carlo methods developed in [2]. Using these, we have investigated the quench protocol described in the main text. We find that rapidly reducing the temperature from the high–temperature paramagnet, to $T < T^* \approx 0.4 \, J_2$, nucleates a large number of vortices. Individual pairs of vortices are observed to spiral towards one another and annihilate in the same way as in the zero–temperature simulations described above. These finite-temperature results will be presented elsewhere.

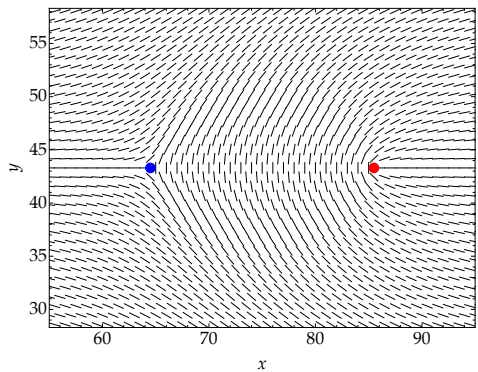

Figure S1. Initial configuration of directors within simulation of pair of vortices shown in the Second Animation [Supplemental Materials] and in Fig. 3 of the main text. Bars show the orientation of the directors, Eq. (S14), with vortices at $\boldsymbol{r}_1$ and $\boldsymbol{r}_2$ labeled with red and blue points, respectively.



\end{document}